\documentclass[iop,deluxetable]{emulateapj}
\usepackage{amsmath,mathptmx}
\newcommand{\Geokerr}{\texttt{Geokerr}}
\newcommand{\Ray}{\texttt{Ray}}
\newcommand{\GRay}{\texttt{GRay}}
\newcommand{\ma}{\smash{\tablenotemark{a}}}
\newcommand{\mb}{\smash{\tablenotemark{b}}}
\newcommand{\mc}{\smash{\tablenotemark{c}}}

\begin{document}

\title{\GRay: a Massively Parallel GPU-Based Code for Ray Tracing in
  Relativistic Spacetimes}

\author{Chi-kwan Chan\altaffilmark{1}, Dimitrios
  Psaltis\altaffilmark{2}, and Feryal \"Ozel\altaffilmark{3,2}}

\affil{Department of Astronomy, University of Arizona,
  933 N. Cherry Ave., Tucson, AZ 85721}
\altaffiltext{1}{NORDITA, Roslagstullsbacken 23, SE-106~91 Stockholm,
  Sweden}
\altaffiltext{2}{Institute for Theory and Computation, Harvard-Smithsonian
  Center for Astrophysics, 60 Garden St., Cambridge, MA 02138}
\altaffiltext{3}{Radcliffe Institute for Advanced Study, Harvard University, 
8 Garden St., Cambridge, MA 02138}

\begin{abstract}
  We introduce \GRay, a massively parallel integrator designed to
  trace the trajectories of billions of photons in a curved spacetime.
  This GPU-based integrator employs the \emph{stream processing}
  paradigm, is implemented in CUDA~C/C++, and runs on nVidia graphics
  cards.
  The peak performance of \GRay\ using single precision floating-point
  arithmetic on a single GPU exceeds 300 GFLOP (or 1~nanosecond per
  photon per time step).
  For a realistic problem, where the peak performance cannot be
  reached, \GRay\ is two orders of magnitude faster than existing
  CPU-based ray tracing codes.
  This performance enhancement allows more effective searches of large
  parameter spaces when comparing theoretical predictions of images,
  spectra, and lightcurves from the vicinities of compact objects to
  observations.
  \GRay\ can also perform on-the-fly ray tracing within general
  relativistic magnetohydrodynamic algorithms that simulate accretion
  flows around compact objects.
  Making use of this algorithm, we calculate the properties of the
  shadows of Kerr black holes and the photon rings that surround them.
  We also provide accurate fitting formulae of their dependencies on
  black hole spin and observer inclination, which can be used to
  interpret upcoming observations of the black holes at the center of
  the Milky Way, as well as M87, with the Event Horizon Telescope.
\end{abstract}


\section{Introduction}

The propagation of photons in the curved spacetimes around black holes
and neutron stars determines the appearance of these compact objects
to an observer at infinity as well as the thermodynamic properties of
the accretion flows around them.
This strong-field lensing imprints characteristic signatures of the
spacetimes on the emerging radiation, which have been exploited in
various attempts to infer the properties of the compact objects
themselves.

As an example, special and general relativistic effects broaden
fluorescence lines that originate in the accretion disks and give them
the characteristic, asymmetric and double-peaked profiles that have
been used in inferring black hole spins in active galactic nuclei and
in galactic sources \citep[see][for a review]{2007ARA&A..45..441M}.
In recent years, this approach has provided strong evidence for rapid
spins in black holes such as MCG~6-30-15 \citep{2006ApJ...652.1028B}
and 1H~0707$-$495 \citep{2009Natur.459..540F} and is expected to
mature even further with upcoming observations with Astro-H
\citep{2012SPIE.8443E..1ZT}.

A similar application of strong-field lensing is encountered in
modeling the images from the accretion flows around the black holes in
the center of the Milky Way and M87
\citep[e.g.,][]{2009ApJ...697...45B, 2009ApJ...703L.142D}.
In the near future, such lensing models will be crucial for
interpreting imaging observations of these two sources with the Event
Horizon Telescope \citep{2009astro2010S..68D}.

Finally, strong-field lensing around a spinning neutron star
determines the pulse profile generated from a hot spots on its
surface.
The pulsation amplitude in such a lightcurve depends sensitively on
the compactness of the neutron star \citep{1983ApJ...274..846P}.
For this reason, comparing model to observed pulsation lightcurves has
led to coarse measurements of the neutron-star properties in
rotation-powered \citep[e.g.,][]{2007ApJ...670..668B} and
accretion-powered millisecond pulsars \citep{2008ApJ...672.1119L} and
bursters \citep[e.g.,][]{2001ApJ...546.1098W, 2002ApJ...581..550M}.
This technique shapes the key science goals of two proposed X-ray
missions, ESA's LOFT \citep{2012SPIE.8443E..2DF} and NASA's NICER
\citep{2009astro2010S...6A}.

The general ray tracing problem in a relativistic spacetime has been
addressed by several research groups to date
\citep[e.g.,][]{1975PhRvD..12..323C, 1983ApJ...274..846P,
  1991ApJ...376...90L, 1995CoPhC..88..109S, 1998ApJ...499L..37M,
  2002ApJ...580.1043B, 2004ragt.meet...33D, 2006MNRAS.366L..10B,
  2007ApJ...654..458C, 2009ApJ...696.1616D, 2009ApJS..184..387D,
  2012ApJ...745....1P, 2012ApJ...753..175B} following two general approaches.
In one approach, which is only applicable to the Kerr spacetime of
spinning black holes, several integrals of motion are used to reduce
the order of the differential equations.
In the other approach, which can be used both in the case of black
holes and neutron stars, the second-order geodesic equations are
integrated.

The Kerr metric is of Petrov-type D and, therefore, the Carter
constant $Q$ provides a third integral of motion along the
trajectories of photons, making the first approach possible \citep[see
  discussion in][]{2010ApJ...718..446J}.
Introducing a deviation from the Kerr metric, however, either in order
to model neutron-star spacetimes, which can have different multipole
moments, or to test the no-hair theorem of black holes, does not
necessarily preserve its Petrov-type D and the Carter constant is no
longer conserved along geodesics (actually no such Killing tensor
exists in spacetimes that are not of Petrov-type D).
As a result, ray tracing in a non-Kerr metric requires integrating the
second-order differential equations for individual geodesics.
Our current algorithm based on \citet{2012ApJ...745....1P}, as well as
those of \citet{2006MNRAS.366L..10B}, \citet{2007ApJ...654..458C}, and
\citet{2009ApJS..184..387D}, follow the latter approach, making them
applicable to a wider range of astrophysical settings.

Although the latter algorithms are not limited by assumptions
regarding the spacetimes of the compact objects and reach efficiencies
of $\simeq 10^4$ geodesic integrations per second, they are still not
at the level of efficiency necessary for the applications discussed
earlier.
For example, in order to simulate the X-ray characteristics of the
accretion flow around a black hole we need to calculate images and
spectra from the innermost $\simeq 100$ gravitational radii around the
black hole.
In order to capture fine details \citep[such as those introduced by
  rays that graze the photon orbit and can affect the detailed images
  and iron-line profiles; see e.g.,][]{2010ApJ...718..446J,
  2005MNRAS.359.1217B} we need to resolve the image plane with a grid
spacing of $\le 0.1$~gravitational radii.
As a result, for a single image, we need to trace at least $10^6$
geodesics.
Even at the current best rate of $\simeq 10^4$ geodesic integrations
per second, a single monoenergetic image at a single instant in time
will require $\sim 100$~seconds on a fast workstation.
This is prohibitively slow, if, for example, we aim to simulate
time-variable emission from a numerical simulation or perform large
parameter studies of black hole spins, accretion rates, and observer
inclinations, when fitting line profiles to data.

A potential resolution to this bottleneck is calculating a large
library of geodesics, storing them on the disk, and using them with an
appropriate interpolation routine either in numerical simulations or
when fitting data.
To estimate the requirements for this approach, we consider, for the
sake of the argument, a rather coarse grid on the image plane,
spanning 100~$M$ in each direction, with a resolution of $1M$.
In principle, we can refine this grid only for those impact parameters
that correspond to geodesics that graze the photon orbit.
In order, e.g., to integrate the radiative transfer equation for each
one of these $10^4$ geodesics that reach the image plane, we need to
store enough information to reproduce the trajectory without
recalculating it.
Assuming a coarse resolution again, we may choose to store $\sim 100$
points per geodesic within the inner 100 gravitational radii.
Along each point, we will need to store at least three components of
the photon 4-momentum (since we can always calculate the fourth
component by the requirement that the photon traces a null geodesic).
For single precision storage (i.e., 4 bytes per number), we will need
to store $4\times 3\times 100\times 10^4=12$~MB of information per
image.
If we want now to use a rather coarse grid of $\sim 30$ values in
black hole spin and $\sim 30$ values in the inclination of the
observer, we need to make use of a $30\times 30\times 12$~MB$=11$~GB
database.
Such a database can only be stored in a hard disk.
At an average latency time of $\simeq 1$~ms for current disks, the
efficiency of this approach cannot exceed $\sim 10^3$ geodesics per
second (given that a typical disk sector has a size of at most 4KB and
can handle the data of no more than a few geodesics).
This is actually comparable to and, in fact, lower than the efficiency
one would achieve by calculating the geodesics in the first place.
Note also, that this estimate was performed for a very coarse grid.

The good news is that ray tracing in vacuum is a trivially
parallelizable algorithm, as individual rays follow independent paths
in the spacetime.
Our goal in this paper is to present a new, massively parallel
algorithm that exploits the recent advances in state-of-the-art
Graphics Processing Unit (GPU) platforms designed specifically to
handle a large number of parallel threads for ray tracing in general
computer visualization (see Figure~\ref{fig:GRay}).

Our algorithm is based on the ray tracing approach of
\citet{2012ApJ...745....1P} and \citet{2012ApJ...753..175B}, employs
nVidia's proprietary Compute Unified Device Architecture (CUDA)
framework, and is implemented in CUDA~C/C++.
We briefly describe the implementation and list the benchmark results
in the next section.
As an application, we take advantage of the speed of the code and
compute the shadows of black holes of different spins at different
inclinations in section~\ref{sec:rings}.
Finally, we discuss future applications of the code such as ray
tracing on-the-fly with general relativistic magnetohydrodynamic
models of accretion flows in section~\ref{sec:discussions}.

\section{Implementation, Numerical Scheme, and Features}
\label{sec:implementation}

\begin{figure}
  \includegraphics[width=\columnwidth]{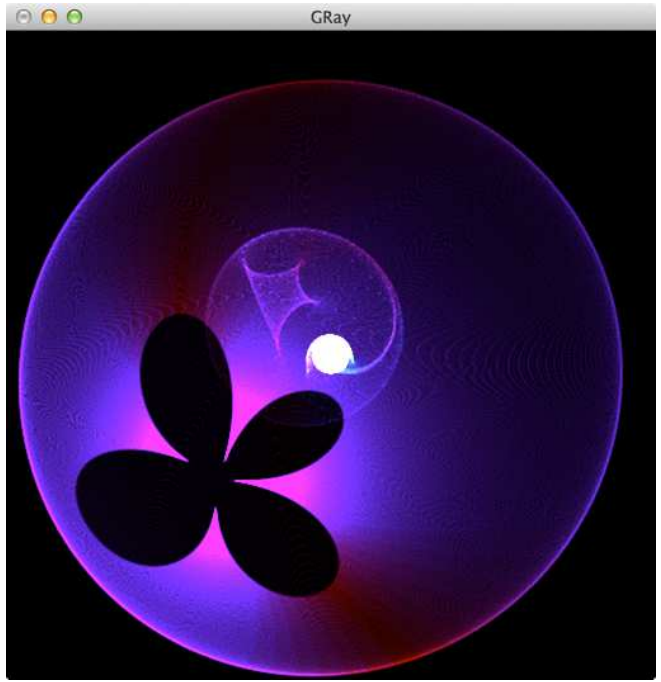}
  \caption{A screen shot of \GRay\ in the interactive mode.
    The image shows the deformation of a plane-parallel grid of
    photons originating at a large distance from a spin $0.999$ black
    hole, as they pass the event horizon (white sphere near the center
    of the screen shot).
    The data always reside on the graphics card memory and the
    visualization is done by CUDA-OpenGL interoperability (see
    section~\ref{sec:implementation}).
    While the geodesic equations are integrated by CUDA, the
    coordinate transformation and particle rendering of the same data
    are both done by the OpenGL shader.}
  \label{fig:GRay}
  \vspace{6pt}
\end{figure}

GPUs were originally developed to handle computationally intensive
graphic applications.
They provide hardware accelerated rendering in computer graphics,
computer-aided design, video games, etc.
Indeed, modern GPUs are optimized specifically with ray tracing in
mind (albeit for what we would call, flat, Euclidean spaces).
However, they have recently found extensive use in scientific
computing, known as General-Purpose (computing on) Graphics Processing
Units (GPGPU, see \texttt{http://gpgpu.org}), as they provide a
low-cost, massively parallel platform for computations that do not
have large memory needs.
These two attributes make GPU technology optimal for the solution of
ray tracing in curved spacetimes.

\begin{figure*}
  \includegraphics[width=\textwidth]{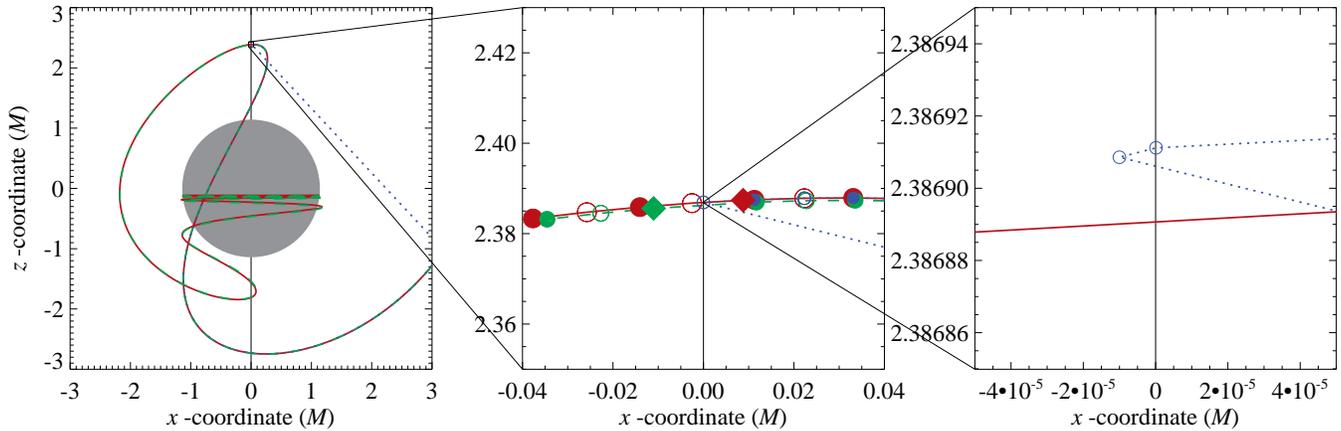} \caption{Numerical
    difficulties occur when the substeps of a 4th-order Runge-Kutta
    update is evaluated very close to, or on the two different sides,
    of the poles.
    In the left panel, the gray circle marks the location of the event
    horizon for a spin 0.99 black hole.
    The vertical black line is the coordinate pole.
    The red solid, green dashed, and blue dotted lines are numerical
    trajectories of the photons with the same initial conditions (see
    text) but with different treatments of the coordinate singularity
    at the pole.
    All three trajectories go around the south pole without any
    apparent problems and wind back to the north pole.
    However, while the red and green trajectories go through the north
    pole and eventually hit the event horizon, the blue trajectory is
    kicked back to infinity.
    The central panel zooms in by a factor of 100 to show that the
    blue trajectory fails to step across the pole.
    To understand this numerical problem, we overplot all the sub- and
    full-steps by open and filled circles, respectively.
    The two overlapping open blue circles are evaluated very close to
    the pole.
    The right panel zooms further in by another factor of 1000.
    It is now clear that the two open blue circles are located on
    different sides of the pole.
    The low-order truncation errors in the 4th-order Runge-Kutta
    scheme, instead of canceling, are enhanced.
    The green trajectory avoids this numerical problem by falling back
    to a 1st-order forward Euler scheme, marked by the green diamond
    in the central panel, whenever the geodesic moves across the pole.
    This 1st-order treatment has a larger truncation error because of
    the 1st-order stepping --- this is visible in the central panel;
    and it may fail if a full step (filled circle) gets too close to the pole.
    To reduce the truncation error and make the integrator more
    robust, the red trajectory uses the forward Euler scheme with a
    smaller time step, which is marked by the red diamond in the
    central panel.
    The subsequent steps are all shifted to avoid the pole.
    This final treatment is what we employ in the production scheme.}
  \label{fig:pole}
  \vspace{6pt}
\end{figure*}

GPUs achieve their high performance by adopting the \emph{stream
  processing} paradigm, which is one kind of single instruction,
multiple data (SIMD) architectures.
There are hundreds\footnote{For example, there are 16 multiprocessors
  on nVidia Tesla M2090.
  Each multiprocessor is made up by 32 cores.
  Hence, there are a total of 512 stream processors on a single GPU.}
of \emph{stream processors} on a single chip.
These stream processors are designed to perform relatively simple
computation in parallel.
On the other hand, the on-chip support of caching (fast memory and
their automatic management) and branching (conditional code execution,
i.e., if-else statements) is primitive.
The developers are responsible for ensuring efficient memory access.

This architecture allows most of the transistors to be devoted to
performance arithmetic and yields an impressive peak performance.
In addition, GPUs hide memory latency by fast switching between
computing threads --- developers are encouraged to over-subscribe the
physical stream processors in order to keep the GPU busy.
This is a very different design compared to general purpose multi-core
Central Processing Units (CPUs), which uses multiple instruction,
multiple data (MIMD) architecture and have intelligent cache
management and branch predictor to maximize the performance.

Although Open Computing Language (OpenCL) is the industrial open
standard of GPGPU programming, we choose CUDA~C/C++ to implement this
publicly available version of \GRay\ because of the availability of
good textbooks \citep[e.g.,][]{Kirk2010, Sanders2010} and the easier
learning curve.
In CUDA terminology, the \emph{host} (i.e., the CPU and the main
memory) sends a parallel task to the \emph{device} (i.e., the GPU and
the graphics card memory) by launching a computing \emph{kernel}.
The kernel runs concurrently on the device involving many lightweight
computing \emph{threads}.
Because of hardware limitation, threads are organized in \emph{blocks}
(of threads) and \emph{grids} (of blocks).
Threads within a block can communicate with each other by using a
small amount of fast, on-chip \emph{shared memory}; while threads
across different blocks can only communicate by accessing a slow,
on-card \emph{global memory}.

Because geodesics do not interact with each other, in \GRay, we simply
put each geodesic into a CUDA thread.
The states of the photons are stored as an array of structure, which,
unfortunately, is not optimal for the GPU to access.
In order to maximize the bandwidth, we employ an \emph{in-block data
  transpose} by using the shared memory\footnote{\GRay\ integrates
  each geodesic for many steps in a single data load.
  The in-block transpose, therefore, only improves \GRay's performance
  in the interactive mode, where we limit the number of steps to trade
  for response time.}.
We fix the block size, i.e., the number of threads within a block,
$n_\mathrm{block}$, to 64, which is larger than the number of physical
stream processors in a multiprocessor.
This over-subscription keeps the GPU busy by allowing a stream
processor to work on a thread while waiting for the data for another
thread to arrive\footnote{Because ray tracing in the Kerr spacetime is
  computationally intensive, this over-subscription does not play a
  crucial role in \GRay's performance.}.
The grid size, i.e., the number of blocks within a grid,
$n_\mathrm{grid}$ is computed by the idiomatic formula \citep[see,
  e.g.,][or sample codes provided by the CUDA software development
  kit]{Kirk2010, Sanders2010}
\begin{align}
  n_\mathrm{grid} = \lfloor(n - 1) / n_\mathrm{block}\rfloor + 1,
\end{align}
where $n$ is the total number of photons and $\lfloor\,\cdot\,\rfloor$
is the floor function.
The above formula ensures that $n_\mathrm{grid} n_\mathrm{block} \ge
n$ so there are enough threads to integrate all the photons.

We employ a standard 4th-order Runge-Kutta scheme presented in
\citet{2012ApJ...745....1P} to integrate equations~(9)--(12).
To avoid the coordinate singularity of the Kerr metric at the event
horizon $r_\mathrm{bh} \equiv 1 + \sqrt{1 - a^2}$, we set the step
size as
\begin{align}
  \Delta\lambda' \equiv \min\left(
  \frac{\Delta}{|d\ln r/d\lambda'| + |d\theta/d\lambda'| + |d\phi/d\lambda'|},
  \frac{r - r_\mathrm{bh}}{2 |dr/d\lambda'|}
  \right),
\end{align}
and stop integrating the photon trajectory at $r_\mathrm{bh} + \delta$
to avoid it crossing the horizon at $r_\mathrm{bh}$.
Both $\Delta \sim 1 / 32$ and $\delta \sim 10^{-6}$ are user provided
parameters.
In addition, we use the remapping
\begin{align}
  \theta, \phi, k_\theta \mapsto
  \begin{cases}
              2\pi -\theta, \phi+\pi, -k_\theta & \mbox{if }\theta > \pi, \\
    \hphantom{2\pi}-\theta, \phi-\pi, -k_\theta & \mbox{if }\theta < 0;
  \end{cases}
  \label{eq:remap}
\end{align}
to enforce $\theta$ to stay in the domain $[0, \pi]$.

The scheme described above can accurately integrate almost all
geodesics.
However, it breaks down for some of the geodesics that pass through
the poles at $\theta = 0$ or $\pi$.
To illustrate how the scheme breaks down, we choose the special
initial conditions $r_0 \cos\theta_0 = 1000 M$, $r_0 \sin\theta_0 =
4.833605 M$, and $\phi_0 = 0$ for which the photon trajectory passes
both the south and north poles of a spin 0.99 black
hole\footnote{Because of cylindrical symmetry, the value of $\phi_0$
  is not important in this setup.
  Indeed, for $\phi_0 = 0^\circ, 1^\circ, 2^\circ, \dots, 359^\circ$,
  the same pole problem is always encountered.}.
In each panel of Figure~\ref{fig:pole}, we plot the result of tracing
the above ray with blue dotted lines.

In the left panel of Figure~\ref{fig:pole}, the gray circle marks the
location of the event horizon for the spin 0.99 black hole.
The vertical black line is the pole.
The green dashed and the red solid lines are the numerical
trajectories of the photons with the same initial conditions but with
different treatments of the coordinate singularity at the pole, as we
will describe below.
All three trajectories go around the south pole without any apparent
problem and wind back to the north pole.
While the red and green trajectories go through the north pole,
circulate around the black hole a couple times, and eventually hit the
event horizon, the blue trajectory is kicked back to infinity due to a
numerical error.

The central panel of Figure~\ref{fig:pole} is a
100$\times$~magnification of the region where the trajectories
intersect with the north pole.
It shows that the blue trajectory fails to step correctly across the
pole.
To pinpoint this numerical difficulty, we overplot all the Runge-Kutta
sub- and full-steps by open and filled circles, respectively.
The two overlapping open blue circles land very close to the pole.

The right panel offers a further 1000$\times$~magnification of the
same region.
It is now clear that the two nearly overlapping open blue circles sit
actually on the opposite sides of the pole.
This is a problem for the 4th-order Runge-Kutta scheme, in which the
solution is assumed to be smooth and can be Taylor expanded.
In this scheme, the low-order truncation errors are normally canceled
by a clever combination of the substeps.
Evaluating the geodesic equation in the different substeps on the two
sides of the pole, however, introduces an inconsistency in the scheme
and enhances the low-order truncation errors.

The green trajectory in Figure~\ref{fig:pole} shows the result of an
improved scheme, which solves the inconsistency by falling back to a
1st-order forward Euler step whenever a geodesic moves across the
pole.
The low-order step is marked by the green diamond in the central
panel.
This treatment mends the numerical difficulty and allows the photon to
pass through the pole.
Unfortunately, the low order stepping results in a larger truncation
error in the numerical solution.
The small but visible offset between the green trajectory and the
other two trajectories in the central panel is indeed caused by the
low order step at the south pole.
Even worse, this treatment may fail when a full step (i.e., the filled
circles) gets too close to the pole.

To reduce the truncation error of the low order step and make the
integrator more robust, in the production scheme of \GRay, we follow
\citet{2012ApJ...745....1P} to monitor the quantity
\begin{align}
  \xi \equiv & \Bigg[
    g_{rr          }\left(\frac{dr     }{d\lambda'}\right)^2 +
    g_{\phi\phi    }\left(\frac{d\phi  }{d\lambda'}\right)^2 +
    g_{\theta\theta}\left(\frac{d\theta}{d\lambda'}\right)^2 \nonumber\\
  & \ \ \ + 
   2g_{t\phi       }\left(\frac{dt     }{d\lambda'}\right)
                    \left(\frac{d\phi  }{d\lambda'}\right)
  \Bigg] \Bigg/ \left[
    g_{tt          }\left(\frac{dt     }{d\lambda'}\right)^2
  \right],
  \label{eq:xi}
\end{align}
which should always remain equal to $-1$.
If $|\xi + 1| > \epsilon$, for some small parameter $\epsilon \sim
10^{-3}$ in the numerical scheme, we \emph{re-integrate} the
inaccurate step by falling back to the 1st-order forward Euler scheme
with a \emph{smaller} time step $\Delta\lambda'/9$.
This step size is chosen so that (i) the absolute numerical error of
the solution does not increase substantially because of this single
low-order step, and (ii) the pole is not encountered even if the Euler
scheme is continuously applied.
This 1st-order step is marked by the red diamond in the central panel
of Figure~\ref{fig:pole}.
The subsequent steps, as shown in the figure by the red circles, are
all shifted towards the left and skip the pole.

We find this final pole treatment extremely robust and use it for all
our production calculations.
For the $3\times10^8$ trajectories that we will integrate in
section~\ref{sec:rings}, none of them fails at the pole as long as we
fix $\Delta = 10^{-6}$.
The rest of the implementation of the algorithm, the initial
conditions, and the setup of the rays on the image plane proceed as in
\citet{2012ApJ...745....1P}.

In addition to performing the computation of ray tracing, \GRay\ takes
advantage of the programmable graphics pipeline to perform real time
data visualization.
It can be compiled in an interactive mode by enabling OpenGL.
The OpenGL frame buffer is allocated on the graphics card, which is
then mapped to CUDA for ray tracing.
This technique is called \emph{CUDA-OpenGL interoperability} --- there
is no need to transfer the data between the host and the device.
Because the data reside on the graphics card and are accessible to
OpenGL, we use the OpenGL Shading Language (GLSL, see
\url{http://www.opengl.org}) to perform coordinate transformation and
sprite drawing.
A screen shot of this built-in real time visualization is provided in
Figure~\ref{fig:GRay}.

\section{Benchmarks}

The theoretical peak performance of a high-end GPU is always about an
order magnitude faster than the peak performance of a high-end
multi-core CPU \citep{Kirk2010, Sanders2010}.
However, because of the fundamental difference in the hardware design,
their real world performances depend on the nature of the problem and
the implementation of the algorithms.
In order to compare different aspects of the implementation of the
ray-tracing algorithm, we perform two different benchmarks on three
codes in this section.
\begin{enumerate}
\item \Geokerr\ is a well established, publicly available code written
  in \texttt{FORTRAN}.
  The code uses a semi-analytical approach to solve for null geodesics
  in Kerr spacetimes, which leads to accurate solutions even with
  arbitrarily large time steps\footnote{Note, however, that substeps
    are need if there are turning points in the null geodesics.}.
  The details of the algorithm are documented in
  \citet{2009ApJ...696.1616D}.
\item \Ray\ is an algorithm that uses a standard 4th-order Runge-Kutta
  scheme to integrate the geodesic equations in spacetimes with
  arbitrary quadrupole moments.
  It is written in C and runs efficiently on CPUs.
  The code has been used to test the no-hair theorem and generate
  profiles and spectra from spinning neutron stars
  \citep{2012ApJ...745....1P, 2012ApJ...753..175B}.
\item \GRay, the open source GPU code we describe in this paper, is
  based on \Ray's algorithm.
  It is written in CUDA~C/C++ and runs efficiently on most nVidia
  GPUs.
  The source code is published under the GNU General Public License
  Version 3 and is available at
  \texttt{\url{https://github.com/chanchikwan/gray}}\,.
\end{enumerate}

\begin{figure}
  \includegraphics[width=\columnwidth,trim=0 0 0 0]{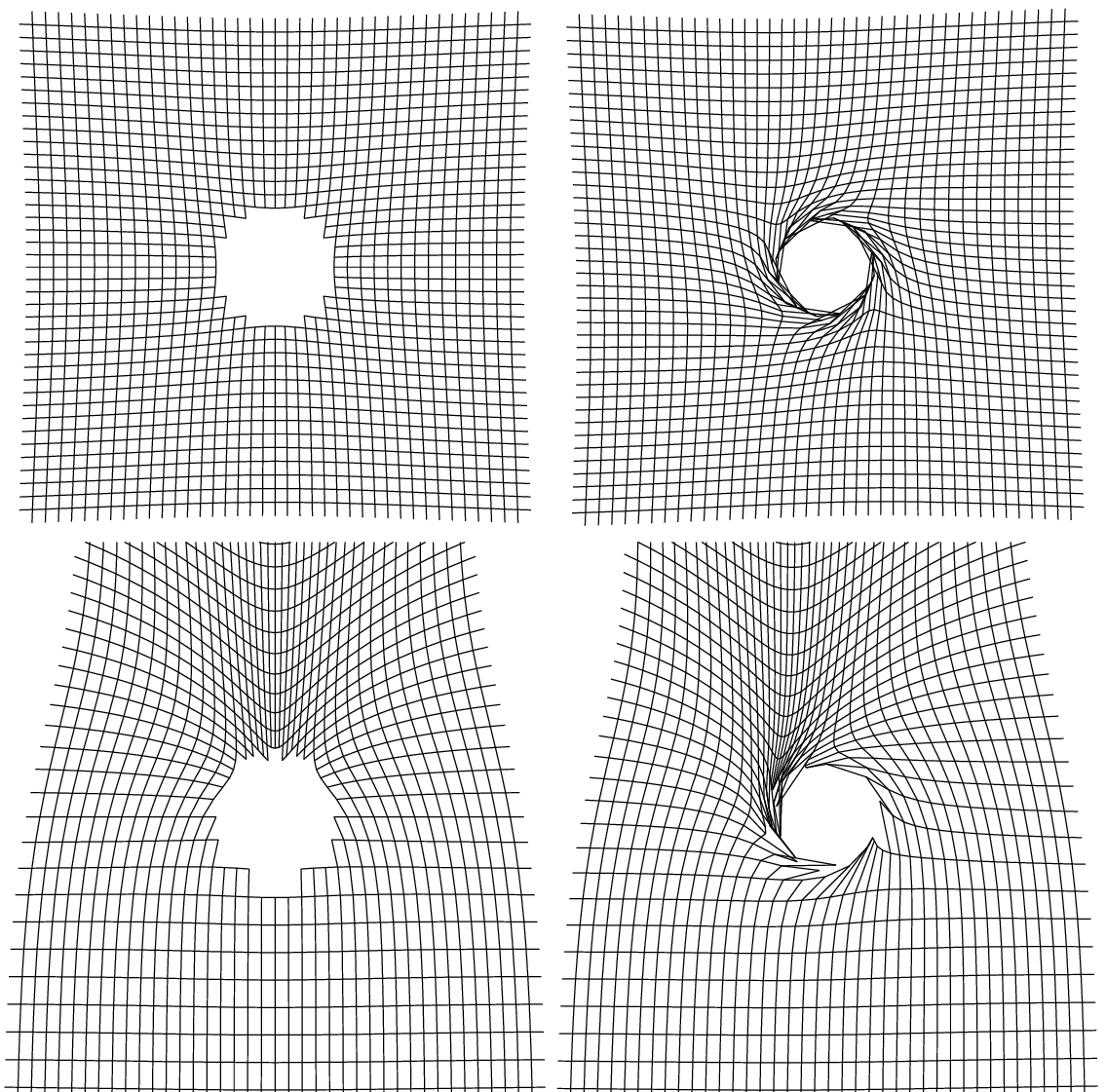}
  \caption{Projections of a uniform Cartesian grid in the image plane
    to the equatorial plane of spin 0 (left column) and 0.95 (right
    column) black holes.
    The images in the top and bottom rows have inclination angles
    $0^\circ$ and $60^\circ$, respectively.
    They are plotted in a way to match Figure~2 of
    \citet{2004ApJ...606.1098S} and Figure~3 of
    \citet{2009ApJ...696.1616D}; i.e., the horizontal and vertical
    axes correspond to the $-\beta_0$- and $-\alpha_0$-directions.
    The configuration in the lower right panel with parameters $a =
    0.95$ and $i = 60^\circ$ is the representative ray tracing problem
    we use in Fig.~\ref{fig:benchmarks} for the comparative benchmarks.}
  \label{fig:projections}
  \vspace{6pt}
\end{figure}

For the first benchmark, we compute the projection of a uniform
Cartesian grid in the image plane onto the equatorial plane of a
spinning black hole.
This problem was carried out in \citet{2004ApJ...606.1098S} and then
used as a test case in \citet{2009ApJ...696.1616D}.
We reproduce the published results in Figure~\ref{fig:projections},
using \GRay\ and initializing the image plane at $r = 1000M$.
The left and right columns show the projections for two black holes
with spins 0 and 0.95, respectively; in each case, the top and bottom
rows represent observer inclinations of $0^\circ$ and $60^\circ$,
respectively.

\begin{figure}
  \includegraphics[width=\columnwidth,trim=0 0 0 0]{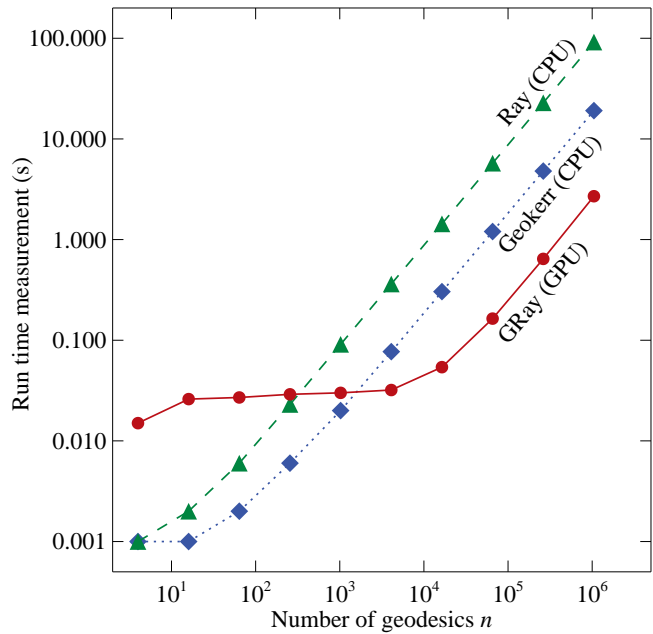}
  \caption{Results of the grid projection benchmark for the
    configuration shown in the lower-right panel of
    Figure~\ref{fig:projections}.
    The run times of three different algorithms, \Geokerr\ (blue
    diamonds), \Ray\ (green triangles), and \GRay\ (red circles) in
    double precision, are plotted against the number of geodesics
    traced for each image.
    The asymptotic linear dependence seen for all three algorithms
    demonstrates explicitly that the ray tracing problem is highly
    parallelizable.
    For a small number of geodesics, the performance of
    \GRay\ flattens to a constant value (approximately 20~ms for the
    configuration used) because of the time required for launching the
    CUDA kernel.}
  \label{fig:benchmarks}
  \vspace{6pt}
\end{figure}

The case shown in the lower-right panel of
Figure~\ref{fig:projections} with parameters $a = 0.95$ and $i =
60^\circ$ is a representative problem, which we will use as a
benchmark.
We use the three algorithms \Geokerr, \Ray, and \GRay\ and calculate
the projection using a grid of $n$ geodesics for each method.
We plot the run time on a single processor of each calculation as a
function of the number of geodesics traced in
Figure~\ref{fig:projections}.

\begin{deluxetable*}{lrrrr}
  \tablewidth{\textwidth}
  \tablecaption{Benchmark results of \GRay\ in comparison to other
    general relativistic ray tracing codes.}
  \tablehead{ \colhead{Processor} & \colhead{\Geokerr\ma} &
    \colhead{\Ray} & \colhead{\GRay} }
  \startdata
  nVidia Tesla   M2090\mb         & \nodata & \nodata &    1.15 \\
nVidia GeForce GT 650M\mb       & \nodata & \nodata &    3.27 \\
nVidia Tesla   M2090            & \nodata & \nodata &    7.15 \\
nVidia GeForce GT 650M          & \nodata & \nodata &   71.87 \\
Intel Core i7-3720QM 2.60GHz\mc &   23000 &  356.67 & \nodata \\
Intel Xeon E5520 2.27GHz\mc     &   43800 &  692.68 & \nodata

  \enddata
  \tablecomments{We focus only at the performance of the geodesic
    integrators.
    The numbers listed in the above table have unit of
    \emph{nanosecond per time step per photon}.
    Hence, smaller number indicates higher performance.}
  \tablenotetext{a}{\Geokerr\ computes the geodesic semi-analytically
    and hence can take arbitrary long time steps unless there is a
    turning point in the geodesic.}
  \tablenotetext{b}{Single precision floating arithmetic is used.}
  \tablenotetext{c}{Both \Geokerr\ and \Ray\ are serial codes.
    Hence, only one CPU core is used in these measurements.}
  \label{tab:benchmarks}
  \vspace{6pt}
\end{deluxetable*}

We can draw a few interesting conclusions from this simple benchmark.
The performance of all algorithms scales linearly for almost all
problems, signifying the fact that ray tracing is a highly
parallelizable problem.
For a small number of rays, the performance of \GRay\ flattens at
about 20~ms for the configuration used, because of the time required
for launching the CUDA Kernel.
This is independent of the number of geodesics but, of course, depends
on the specific hardware, drivers, and operating system used.
For a MacBook~Pro running OS~X, this time is of order a few tens of
milliseconds, while the launching time may be as large as 0.5 seconds
for some Linux configurations.

For calculations with a large number of geodesics, which is the regime
that motivated our work, \GRay\ is faster than both \Geokerr\ and
\Ray\ by one to two orders of magnitude.
It is important to emphasize that the performance of \GRay\ exceeds
that of the other algorithms even in this benchmark that is designed
in a way that favors the semi-analytical approach of \Geokerr.
This is true because we are only interested in the intersection of the
ray with the equatorial plane, which \Geokerr\ can achieve with a very
small number of steps per ray.
In more general radiative transfer problems, however, we have to
divide each ray in small steps in all methods in order to integrate
accurately the radiative transfer equation through black hole
accretion flows.
This requirement puts the Runge-Kutta integrators at a larger
advantage compared to semi-analytic approaches.

In order to assess the performance of \GRay\ in this second situation,
we setup a benchmark to measure the average time that the integrators
require to take a single step in the integration of a photon path.
We list the results of this benchmark for the three algorithms in
Table~\ref{tab:benchmarks}, where the numbers have unit of
\emph{nanoseconds per timestep per photon}, such that a smaller number
indicates higher performance.
In this benchmark, the benefit of the GPU integrator becomes clearly
visible as \GRay\ is 50 times faster than \Ray\ and more than a factor
of 1000 faster than \Geokerr.

\section{Properties of Photon Rings around Kerr Black Holes}
\label{sec:rings}

\begin{figure*}
  \includegraphics[width=\textwidth]{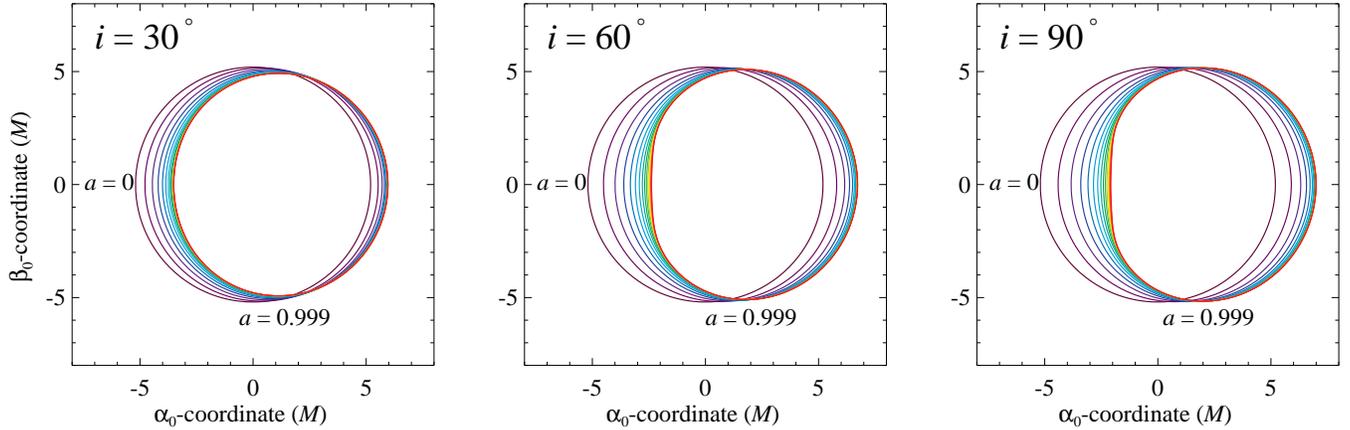}
  \caption{Photon rings around Kerr black holes with different spins
    $a$ and observer inclinations $i$.
    The left, central, and right panels show the photon rings for $i =
    30^\circ$, $60^\circ$, and $90^\circ$, respectively.
    In each panel, different colors represent different spins --- from
    black being $a = 0$ to red being $a = 0.999$.
    For each inclination, the size of the photon ring depends very
    weakly on the black hole spin.
    Moreover, the photon ring retains its nearly circular shape even
    at high black hole spins; a significant distortion appears only
    for $a \gtrsim 0.99$ and at large inclination angles.}
  \label{fig:shadow(a)}
  \vspace{6pt}
\end{figure*}

\begin{figure*}
  \includegraphics[width=\textwidth]{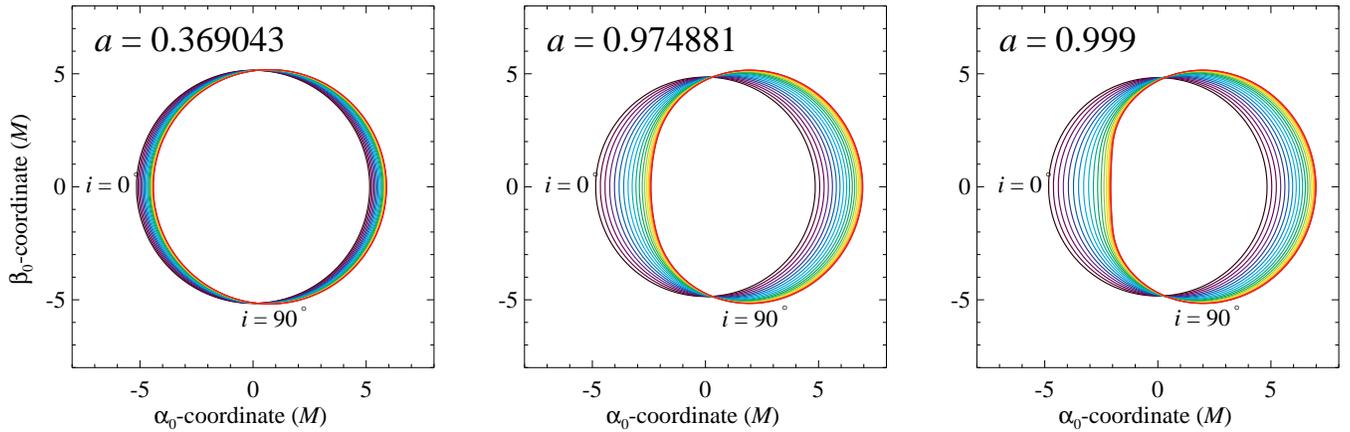} \caption{Photon rings
    around Kerr black holes with different spins $a$ and observer
    inclinations $i$.
    The left, central, and right panels plot the photon rings for $a =
    0.369043$, $0.974881$, and $0.999$, respectively.
    In each panel, different colors represent different inclinations
    --- going from black for $i = 0^\circ$, to blue for $i = 5^\circ$,
    $10^\circ$, \dots, to red for $i = 90^\circ$.
    The photon rings become asymmetric only for $a \gtrsim 0.99$ and
    at large inclination angles.}
  \label{fig:shadow(i)}
  \vspace{6pt}
\end{figure*}

\begin{figure*}
  \includegraphics[width=\textwidth]{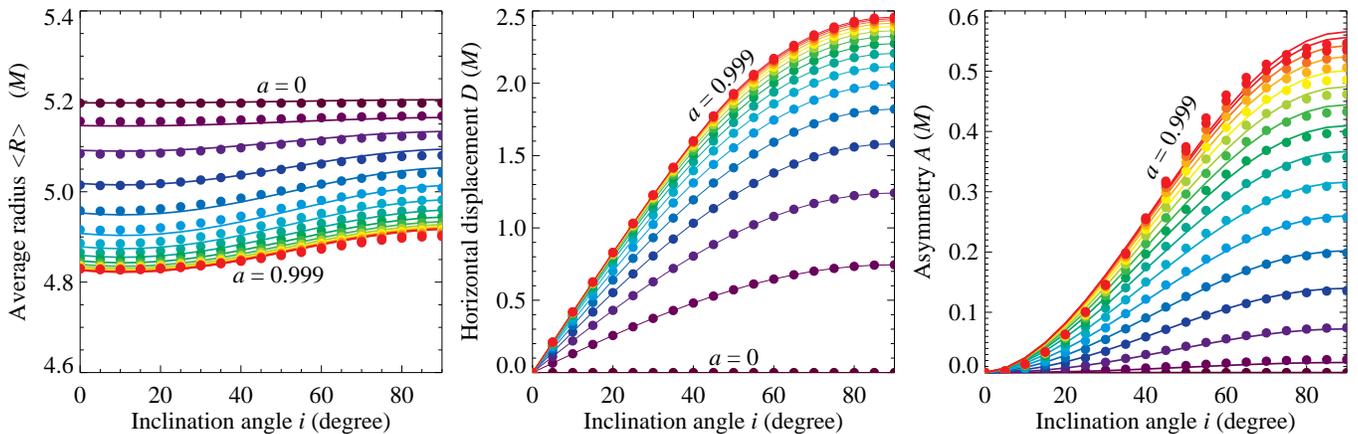}
  \caption{Photon ring properties for Kerr black holes with different
    spins $a$ and different inclinations $i$.
    The left, central, and right panels show the average radius
    $\langle R\rangle$, the horizontal displacement $D$, and the
    asymmetry parameter $A$ of the rings, respectively.
    In each panel, the horizontal axis is the inclination $i$ and
    different colors represent different spins --- from black for $a =
    0$ to red for $a = 0.999$.
    In the leftmost and rightmost panels, the solid curves show the
    analytic fits discussed in the text.}
  \label{fig:properties}
  \vspace{6pt}
\end{figure*}

\begin{figure}
  \includegraphics[width=\columnwidth,trim=0 0 0 0]{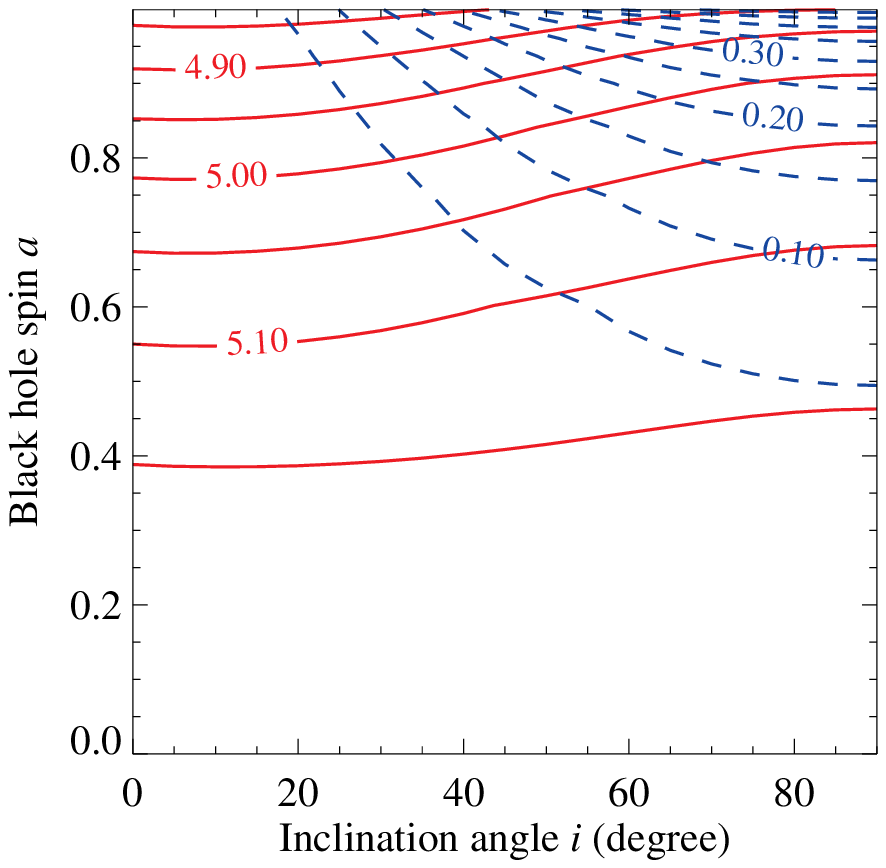}
  \caption{Contours of constant photon ring properties around Kerr
    black holes, which can be inferred with imaging observations of
    their accretion flows.
    The contours of constant average radius of the ring, $\langle
    R\rangle$, and asymmetry parameter, $A$, are plotted as solid red
    and dashed blue curves, respectively.
    If the two contour lines that correspond to an observed
    photon-ring radius and asymmetry for a black hole cross, then both
    the spin of the black hole and the inclination of the observer can
    be independently inferred.
    On the other hand, if the two contour lines do not intersect, this
    will indicate a violation of the no-hair theorem
    \citep{2010ApJ...718..446J}.}
  \label{fig:contours}
  \vspace{6pt}
\end{figure}

Being a massively parallel algorithm, \GRay\ is an ideal tool to study
black hole images that involve integrating billions of photon
trajectories.
In general, the details of black hole images depend on the
time-dependent properties of the turbulent accretion flows (see also
section~\ref{sec:discussions} for a detailed discussion).
In all cases, however, for optically thin accretion flows such as the
one expected around Sgr~A* at mm wavelengths, the projection of the
circular photon orbit produces a bright ring on the image plane that
stands out against the background \citep[see][]{1979A&A....75..228L,
  2005MNRAS.359.1217B, 2010ApJ...718..446J}.
As pointed out by \citet{2010ApJ...718..446J}, the shape of this so
called \emph{photon ring} that surrounds the black hole shadow is a
general relativistic effect and is insensitive to the complicated
astrophysics of the accretion flows.
Careful matching of the theoretical predictions of the photon ring
with observations, therefore, provides an unmistakable way to measure
the black hole mass and even to test the no-hair theorem
\citep{2010ApJ...718..446J, 2012ApJ...758...30J}.

We performed a systematic calculation of the photon rings around Kerr
black holes of different spins $a$ and observer inclinations $i$.
We choose 16 values of spin according to the relation
\begin{align}
  a_j = 1 - 10^{-j/5}, \mbox{ where }j = 0, 1, \dots, 15
\end{align}
so that $1-a$ is evenly spaced in log scale, and 19 values of
inclination $i = 0, 5, 10, \dots, 90$.
For each configuration, we set up the image plane at $r = 1000M$ and
define its center at the intersection of this plane with a radial line
emerging out of the black hole.
We define $(\mathcal{R}, \vartheta)$ to be the local polar coordinate
of the image plane.
We set up a grid of $6000\times181$ rays in the polar domain $(1.5,
7.5)\times[0,\pi]$ and integrate them toward the black hole.
Hence, there are $16 \times 19 \times 6000 \times 181 \approx
3\times10^8$ geodesics in this parameter study\footnote{We use
  \texttt{git} to manage the source code of \GRay.
  For reproducibility, the setup of this parameter study is available
  in the source repository with commit id \texttt{4e20d4c0}.
  See also the commit message for running the study.}.

We plot the outlines of the photon rings in
Figures~\ref{fig:shadow(a)} and \ref{fig:shadow(i)}.
As discussed in \citet{2010ApJ...718..446J}, we find that the size of
the photon ring depends very weakly on the spin of the black hole and
the inclination of the observer.
Moreover, the ring retains a highly circular shape at even high spins,
as significant asymmetries appear only at $a \gtrsim 0.99$.
In \citet{2010ApJ...718..446J}, we attributed this to the cancellation
of the ellipsoidal geometry of the Kerr spacetime by the
frame-dragging effects on the propagation of photons, which appears to
be exact at the quadrupole order.

In order to quantify the magnitude of the effects discussed above, we
follow \citet{2010ApJ...718..446J} to define the horizontal
displacement of the ring from the geometric center of the spacetime as
\begin{align}
  D \equiv \frac{|\alpha_{0,\min} + \alpha_{0,\max}|}{2},
\end{align}
the average radius of the ring as
\begin{align}
  \langle R\rangle \equiv \frac{1}{2\pi}\int_0^{2\pi} R d\vartheta,
\end{align}
where $R \equiv \left[(\alpha_0 - D)^2 + \beta_0^2\right]^{1/2}$ and
$\theta \equiv \tan^{-1}(\beta_0/\alpha_0)$, and the asymmetry
parameter as
\begin{align}
  A \equiv 2 \left[\frac{1}{2\pi}\int_0^{2\pi}
    \left(R - \langle R\rangle\right)^2 d\vartheta\right]^{1/2}.
\end{align}
In the above equations, the coordinates $\alpha_0$ and $\beta_0$ are
understood to be measured on a two-dimensional Cartesian coordinate
system on the image plane, i.e., they are related to $\mathcal{R}$ and
$\vartheta$ by the coordinate transformation
\begin{align}
  \alpha_0 &= \mathcal{R} \cos\vartheta, \\
  \beta_0 &= \mathcal{R} \sin\vartheta.
\end{align}
In Figure~\ref{fig:properties}, we plot these ring quantities as
functions of the observer inclination $i$ at different black hole
spins $a$.

In order to facilitate the comparison of theoretical models to
upcoming observations of black hole shadows with the Event Horizon
Telescope, we have obtained simple analytic fits to the dependence of
the average radius and asymmetry of the photon ring for black holes
with different spins and observer inclinations.
In particular, we find
\begin{align}
  \langle R \rangle \simeq R_0 + R_1 \cos(2.14 i - 22.9^\circ)
  \label{eq:Rapprox}
\end{align}
with
\begin{align}
  R_0 &= \left(5.2 - 0.209 a + 0.445 a^2 - 0.567 a^3\right) M\nonumber\\
  R_1 &= \left[0.24 - \frac{3.3}{(a - 0.9017)^2 + 0.059}\right]\times 10^{-3} M
\end{align}
and
\begin{align}
  A\simeq A_0\sin^{n}i\;,
  \label{eq:Aapprox}
\end{align}
with
\begin{align}
  A_0 &= \left(0.332 a^3 + 0.176 a^{21.7} + 0.0756 a^{195}\right) M\nonumber\\
  n   &= 1.55 (1 - a)^{-0.022} + 1.3 (1 - a)^{0.98}\;.
\end{align}
In all relations, the arguments of the trigonometric functions are in
degrees.
The above empirical relations are shown as solid curves in the
leftmost and rightmost panels of Figure~\ref{fig:properties}.

In Figure~\ref{fig:contours}, we provide a different representation of
the above results, by plotting contours of constant average radius
$\langle R\rangle$ and asymmetry parameter $A$ on the parameter space
of black hole spin $a$ and observer inclination $i$.
The Event Horizon Telescope \citep{2009astro2010S..68D} aims to
perform imaging observations of the inner accretion flows around the
black holes in the center of the Milky Way and of M87, in order to
measure these two parameters of the black hole shadows.
(The displacement $D$ cannot be readily measured, since there is very
little indication of the geometric center of the spacetime that can be
obtained from the images).
If the two contour lines that correspond to the observed radius and
asymmetry for each black hole cross, then both the spin of the black
hole and the inclination of the observer can be inferred independently
using such observations.
On the other hand, if the two contour lines do not intersect, this
will indicate a violation of the no-hair theorem
\citep{2010ApJ...718..446J}.

\section{Discussions}
\label{sec:discussions}

In this paper, we presented our implementation of the massively
parallel ray tracing algorithm \GRay\ for GPU architecture.
We demonstrated that its performance is about two orders of magnitude
faster than equivalent CPU ray tracing codes.
Running this algorithm on an nVidia Tesla M2090 card, we are able to
compute a $1024 \times 1024$ pixel image in about a few seconds (see
Figure~\ref{fig:benchmarks}).
At the same time, we can achieve timesteps per photon as small as 1~ns
on the same GPU card (see Table~\ref{tab:benchmarks}).
Bearing in mind that communication is almost always slower than
computation in high-performance computing (e.g., the host-device
bandwidth, through PCI express, is at least an order of magnitude
faster than the hard disk bandwidth, through SATA) also leads us to
conclude that using GPUs to perform the computation of geodesics when
needed in an algorithm is a more efficient approach to solving this
problem compared to tabulating precomputed results in a database.

Our initial goal is to use \GRay\ to make significant advances in
modeling and interpreting observational data.
Nevertheless, \GRay\ will also be extremely useful in performing three
dimensional, magnetohydrodynamic (MHD) calculations in full general
relativity aiming to achieve ab initio simulations of MHD processes in
the vicinity of black hole horizons \citep[see,
  e.g.,][]{2003ApJ...589..458D, 2003ApJ...589..444G,
  2006astro.ph..9004M, 2007CQGra..24S.235G, 2007A&A...473...11D,
  2008A&A...492..937C, 2011arXiv1102.5202Z}.
Besides being very important for improving of our understanding of
accretion flows, MHD simulations have been instrumental in
interpreting observations of Sgr~A* and its unusual flares \citep[see,
  e.g.,][]{2009ApJ...701..521C, 2009ApJ...706..497M,
  2010ApJ...725..450D, 2012ApJ...746L..10D}.

Comparing the results of numerical simulations to observations
requires, at the very least, using the calculated time-dependent
thermodynamic and hydrodynamic properties of the MHD flows to predict
lightcurves, spectra, and images.
At the same time, the propagation of radiation within the MHD flow
contributes to its heating and cooling.
In addition, radiation forces determine even the dynamics of near
Eddington accretion flows.
Calculating the propagation of radiation within the accretion flow and
to an observer at infinity in a time-dependent manner is very time
consuming.
It has been taken into account only in limited simulations and under
various simplifying assumptions \citep[see,
  e.g.,][]{2008arXiv0802.0848D}.
In fact, only a handful of numerical algorithms have been used to date
for calculations of observed quantities post facto, based on
\emph{snapshots} of MHD simulations \citep{2009ApJ...696.1616D,
  2009ApJS..184..387D}.
This ``fast light'' approximation breaks down close to the black hole
because of the speed of the plasma there is comparable to the speed of
light.
In order to overcome the storage requirement of frequent data dump,
\GRay\ may be integrated into a general relativistic MHD code to
perform ray tracing on-the-fly.

\acknowledgements

This work was supported in part by the NSF grant AST-1108753, NSF
CAREER award AST-0746549, and Chandra Theory grant TM2-13002X. F.\"O.
gratefully acknowledges support from the Radcliffe Institute for
Advanced Study at Harvard University.

\bibliography{ms,my}

\begin{thebibliography}{45}
\expandafter\ifx\csname natexlab\endcsname\relax\def\natexlab#1{#1}\fi

\bibitem[{{Arzoumanian} {et~al.}(2009){Arzoumanian}, {Bogdanov}, {Cordes},
  {Gendreau}, {Lai}, {Lattimer}, {Link}, {Lommen}, {Miller}, {Ray}, {Rutledge},
  {Strohmayer}, {Wilson-Hodge}, \& {Wood}}]{2009astro2010S...6A}
{Arzoumanian}, Z., {Bogdanov}, S., {Cordes}, J., {et~al.} 2009, in Astronomy,
  Vol. 2010, astro2010: The Astronomy and Astrophysics Decadal Survey, 6

\bibitem[{{Baub{\"o}ck} {et~al.}(2012){Baub{\"o}ck}, {Psaltis}, {{\"O}zel}, \&
  {Johannsen}}]{2012ApJ...753..175B}
{Baub{\"o}ck}, M., {Psaltis}, D., {{\"O}zel}, F., \& {Johannsen}, T. 2012,
  \apj, 753, 175

\bibitem[{{Beckwith} \& {Done}(2005)}]{2005MNRAS.359.1217B}
{Beckwith}, K., \& {Done}, C. 2005, \mnras, 359, 1217

\bibitem[{{Bogdanov} {et~al.}(2007){Bogdanov}, {Rybicki}, \&
  {Grindlay}}]{2007ApJ...670..668B}
{Bogdanov}, S., {Rybicki}, G.~B., \& {Grindlay}, J.~E. 2007, \apj, 670, 668

\bibitem[{{Braje} \& {Romani}(2002)}]{2002ApJ...580.1043B}
{Braje}, T.~M., \& {Romani}, R.~W. 2002, \apj, 580, 1043

\bibitem[{{Brenneman} \& {Reynolds}(2006)}]{2006ApJ...652.1028B}
{Brenneman}, L.~W., \& {Reynolds}, C.~S. 2006, \apj, 652, 1028

\bibitem[{{Broderick}(2006)}]{2006MNRAS.366L..10B}
{Broderick}, A.~E. 2006, \mnras, 366, L10

\bibitem[{{Broderick} {et~al.}(2009){Broderick}, {Fish}, {Doeleman}, \&
  {Loeb}}]{2009ApJ...697...45B}
{Broderick}, A.~E., {Fish}, V.~L., {Doeleman}, S.~S., \& {Loeb}, A. 2009, \apj,
  697, 45

\bibitem[{{Cadeau} {et~al.}(2007){Cadeau}, {Morsink}, {Leahy}, \&
  {Campbell}}]{2007ApJ...654..458C}
{Cadeau}, C., {Morsink}, S.~M., {Leahy}, D., \& {Campbell}, S.~S. 2007, \apj,
  654, 458

\bibitem[{{Cerd{\'a}-Dur{\'a}n} {et~al.}(2008){Cerd{\'a}-Dur{\'a}n}, {Font},
  {Ant{\'o}n}, \& {M{\"u}ller}}]{2008A&A...492..937C}
{Cerd{\'a}-Dur{\'a}n}, P., {Font}, J.~A., {Ant{\'o}n}, L., \& {M{\"u}ller}, E.
  2008, \aap, 492, 937

\bibitem[{{Chan} {et~al.}(2009){Chan}, {Liu}, {Fryer}, {Psaltis}, {{\"O}zel},
  {Rockefeller}, \& {Melia}}]{2009ApJ...701..521C}
{Chan}, C.-k., {Liu}, S., {Fryer}, C.~L., {et~al.} 2009, \apj, 701, 521

\bibitem[{{Cunningham}(1975)}]{1975PhRvD..12..323C}
{Cunningham}, C.~T. 1975, \prd, 12, 323

\bibitem[{{De Villiers}(2008)}]{2008arXiv0802.0848D}
{De Villiers}, J.-P. 2008, ArXiv e-prints

\bibitem[{{De Villiers} \& {Hawley}(2003)}]{2003ApJ...589..458D}
{De Villiers}, J.-P., \& {Hawley}, J.~F. 2003, \apj, 589, 458

\bibitem[{{Del Zanna} {et~al.}(2007){Del Zanna}, {Zanotti}, {Bucciantini}, \&
  {Londrillo}}]{2007A&A...473...11D}
{Del Zanna}, L., {Zanotti}, O., {Bucciantini}, N., \& {Londrillo}, P. 2007,
  \aap, 473, 11

\bibitem[{{Dexter} \& {Agol}(2009)}]{2009ApJ...696.1616D}
{Dexter}, J., \& {Agol}, E. 2009, \apj, 696, 1616

\bibitem[{{Dexter} {et~al.}(2009){Dexter}, {Agol}, \&
  {Fragile}}]{2009ApJ...703L.142D}
{Dexter}, J., {Agol}, E., \& {Fragile}, P.~C. 2009, \apjl, 703, L142

\bibitem[{{Dodds-Eden} {et~al.}(2010){Dodds-Eden}, {Sharma}, {Quataert},
  {Genzel}, {Gillessen}, {Eisenhauer}, \& {Porquet}}]{2010ApJ...725..450D}
{Dodds-Eden}, K., {Sharma}, P., {Quataert}, E., {et~al.} 2010, \apj, 725, 450

\bibitem[{{Doeleman} {et~al.}(2009){Doeleman}, {Agol}, {Backer}, {Baganoff},
  {Bower}, {Broderick}, {Fabian}, {Fish}, {Gammie}, {Ho}, {Honman},
  {Krichbaum}, {Loeb}, {Marrone}, {Reid}, {Rogers}, {Shapiro}, {Strittmatter},
  {Tilanus}, {Weintroub}, {Whitney}, {Wright}, \&
  {Ziurys}}]{2009astro2010S..68D}
{Doeleman}, S., {Agol}, E., {Backer}, D., {et~al.} 2009, in Astronomy, Vol.
  2010, astro2010: The Astronomy and Astrophysics Decadal Survey, 68

\bibitem[{{Dolence} {et~al.}(2009){Dolence}, {Gammie}, {Mo{\'s}cibrodzka}, \&
  {Leung}}]{2009ApJS..184..387D}
{Dolence}, J.~C., {Gammie}, C.~F., {Mo{\'s}cibrodzka}, M., \& {Leung}, P.~K.
  2009, \apjs, 184, 387

\bibitem[{{Dolence} {et~al.}(2012){Dolence}, {Gammie}, {Shiokawa}, \&
  {Noble}}]{2012ApJ...746L..10D}
{Dolence}, J.~C., {Gammie}, C.~F., {Shiokawa}, H., \& {Noble}, S.~C. 2012,
  \apjl, 746, L10

\bibitem[{{Dov{\v c}iak} {et~al.}(2004){Dov{\v c}iak}, {Karas}, {Martocchia},
  {Matt}, \& {Yaqoob}}]{2004ragt.meet...33D}
{Dov{\v c}iak}, M., {Karas}, V., {Martocchia}, A., {Matt}, G., \& {Yaqoob}, T.
  2004, in RAGtime 4/5: Workshops on black holes and neutron stars, ed.
  S.~{Hled{\'{\i}}k} \& Z.~{Stuchl{\'{\i}}k}, 33--73

\bibitem[{{Fabian} {et~al.}(2009){Fabian}, {Zoghbi}, {Ross}, {Uttley}, {Gallo},
  {Brandt}, {Blustin}, {Boller}, {Caballero-Garcia}, {Larsson}, {Miller},
  {Miniutti}, {Ponti}, {Reis}, {Reynolds}, {Tanaka}, \&
  {Young}}]{2009Natur.459..540F}
{Fabian}, A.~C., {Zoghbi}, A., {Ross}, R.~R., {et~al.} 2009, \nat, 459, 540

\bibitem[{{Feroci} {et~al.}(2012){Feroci}, {den Herder}, {Bozzo}, {Barret},
  {Brandt}, {Hernanz}, {van der Klis}, {Pohl}, {Santangelo}, {Stella}, \&
  et~al.}]{2012SPIE.8443E..2DF}
{Feroci}, M., {den Herder}, J.~W., {Bozzo}, E., {et~al.} 2012, in Society of
  Photo-Optical Instrumentation Engineers (SPIE) Conference Series, Vol. 8443,
  Society of Photo-Optical Instrumentation Engineers (SPIE) Conference Series

\bibitem[{{Gammie} {et~al.}(2003){Gammie}, {McKinney}, \&
  {T{\'o}th}}]{2003ApJ...589..444G}
{Gammie}, C.~F., {McKinney}, J.~C., \& {T{\'o}th}, G. 2003, \apj, 589, 444

\bibitem[{{Giacomazzo} \& {Rezzolla}(2007)}]{2007CQGra..24S.235G}
{Giacomazzo}, B., \& {Rezzolla}, L. 2007, Classical and Quantum Gravity, 24,
  235

\bibitem[{{Johannsen} \& {Psaltis}(2010)}]{2010ApJ...718..446J}
{Johannsen}, T., \& {Psaltis}, D. 2010, \apj, 718, 446

\bibitem[{{Johannsen} {et~al.}(2012){Johannsen}, {Psaltis}, {Gillessen},
  {Marrone}, {{\"O}zel}, {Doeleman}, \& {Fish}}]{2012ApJ...758...30J}
{Johannsen}, T., {Psaltis}, D., {Gillessen}, S., {et~al.} 2012, \apj, 758, 30

\bibitem[{Kirk \& Hwu(2010)}]{Kirk2010}
Kirk, D.~B., \& Hwu, W.-m.~W. 2010, Programming Massively Parallel Processors:
  A Hands-on Approach, 1st edn. (San Francisco, CA, USA: Morgan Kaufmann
  Publishers Inc.)

\bibitem[{{Laor}(1991)}]{1991ApJ...376...90L}
{Laor}, A. 1991, \apj, 376, 90

\bibitem[{{Leahy} {et~al.}(2008){Leahy}, {Morsink}, \&
  {Cadeau}}]{2008ApJ...672.1119L}
{Leahy}, D.~A., {Morsink}, S.~M., \& {Cadeau}, C. 2008, \apj, 672, 1119

\bibitem[{{Luminet}(1979)}]{1979A&A....75..228L}
{Luminet}, J.-P. 1979, \aap, 75, 228

\bibitem[{{Miller}(2007)}]{2007ARA&A..45..441M}
{Miller}, J.~M. 2007, \araa, 45, 441

\bibitem[{{Miller} \& {Lamb}(1998)}]{1998ApJ...499L..37M}
{Miller}, M.~C., \& {Lamb}, F.~K. 1998, \apjl, 499, L37

\bibitem[{{Mizuno} {et~al.}(2006){Mizuno}, {Nishikawa}, {Koide}, {Hardee}, \&
  {Fishman}}]{2006astro.ph..9004M}
{Mizuno}, Y., {Nishikawa}, K.-I., {Koide}, S., {Hardee}, P., \& {Fishman},
  G.~J. 2006, ArXiv Astrophysics e-prints

\bibitem[{{Mo{\'s}cibrodzka} {et~al.}(2009){Mo{\'s}cibrodzka}, {Gammie},
  {Dolence}, {Shiokawa}, \& {Leung}}]{2009ApJ...706..497M}
{Mo{\'s}cibrodzka}, M., {Gammie}, C.~F., {Dolence}, J.~C., {Shiokawa}, H., \&
  {Leung}, P.~K. 2009, \apj, 706, 497

\bibitem[{{Muno} {et~al.}(2002){Muno}, {{\"O}zel}, \&
  {Chakrabarty}}]{2002ApJ...581..550M}
{Muno}, M.~P., {{\"O}zel}, F., \& {Chakrabarty}, D. 2002, \apj, 581, 550

\bibitem[{{Pechenick} {et~al.}(1983){Pechenick}, {Ftaclas}, \&
  {Cohen}}]{1983ApJ...274..846P}
{Pechenick}, K.~R., {Ftaclas}, C., \& {Cohen}, J.~M. 1983, \apj, 274, 846

\bibitem[{{Psaltis} \& {Johannsen}(2012)}]{2012ApJ...745....1P}
{Psaltis}, D., \& {Johannsen}, T. 2012, \apj, 745, 1

\bibitem[{Sanders \& Kandrot(2010)}]{Sanders2010}
Sanders, J., \& Kandrot, E. 2010, CUDA by Example: An Introduction to
  General-Purpose GPU Programming, 1st edn. (Addison-Wesley Professional)

\bibitem[{{Schnittman} \& {Bertschinger}(2004)}]{2004ApJ...606.1098S}
{Schnittman}, J.~D., \& {Bertschinger}, E. 2004, \apj, 606, 1098

\bibitem[{{Speith} {et~al.}(1995){Speith}, {Riffert}, \&
  {Ruder}}]{1995CoPhC..88..109S}
{Speith}, R., {Riffert}, H., \& {Ruder}, H. 1995, Computer Physics
  Communications, 88, 109

\bibitem[{{Takahashi} {et~al.}(2012){Takahashi}, {Mitsuda}, {Kelley}, {Aarts},
  {Aharonian}, {Akamatsu}, {Akimoto}, {Allen}, {Anabuki}, {Angelini}, {Arnaud},
  {Asai}, {Audard}, {Awaki}, {Azzarello}, {Baluta}, {Bamba}, {Bando}, {Bautz},
  {Blandford}, {Boyce}, {Brown}, {Cackett}, {Chernyakova}, {Coppi},
  {Costantini}, {de Plaa}, {den Herder}, {DiPirro}, {Done}, {Dotani}, {Doty},
  {Ebisawa}, {Eckart}, {Enoto}, {Ezoe}, {Fabian}, {Ferrigno}, {Foster},
  {Fujimoto}, {Fukazawa}, {Funk}, {Furuzawa}, {Galeazzi}, {Gallo}, {Gandhi},
  {Gendreau}, {Gilmore}, {Haas}, {Haba}, {Hamaguchi}, {Hatsukade}, {Hayashi},
  {Hayashida}, {Hiraga}, {Hirose}, {Hornschemeier}, {Hoshino}, {Hughes},
  {Hwang}, {Iizuka}, {Inoue}, {Ishibashi}, {Ishida}, {Ishimura}, {Ishisaki},
  {Ito}, {Iwata}, {Iyomoto}, {Kaastra}, {Kallman}, {Kamae}, {Kataoka},
  {Katsuda}, {Kawahara}, {Kawaharada}, {Kawai}, {Kawasaki}, {Khangaluyan},
  {Kilbourne}, {Kimura}, {Kinugasa}, {Kitamoto}, {Kitayama}, {Kohmura},
  {Kokubun}, {Kosaka}, {Koujelev}, {Koyama}, {Krimm}, {Kubota}, {Kunieda},
  {LaMassa}, {Laurent}, {Lebrun}, {Leutenegger}, {Limousin}, {Loewenstein},
  {Long}, {Lumb}, {Madejski}, {Maeda}, {Makishima}, {Marchand}, {Markevitch},
  {Matsumoto}, {Matsushita}, {McCammon}, {McNamara}, {Miller}, {Miller},
  {Mineshige}, {Minesugi}, {Mitsuishi}, {Miyazawa}, {Mizuno}, {Mori}, {Mori},
  {Mukai}, {Murakami}, {Murakami}, {Mushotzky}, {Nagano}, {Nagino}, {Nakagawa},
  {Nakajima}, {Nakamori}, {Nakazawa}, {Namba}, {Natsukari}, {Nishioka},
  {Nobukawa}, {Nomachi}, {O'Dell}, {Odaka}, {Ogawa}, {Ogawa}, {Ogi}, {Ohashi},
  {Ohno}, {Ohta}, {Okajima}, {Okamoto}, {Okazaki}, {Ota}, {Ozaki}, {Paerels},
  {Paltani}, {Parmar}, {Petre}, {Pohl}, {Porter}, {Ramsey}, {Reis}, {Reynolds},
  {Russell}, {Safi-Harb}, {Sakai}, {Sameshima}, {Sanders}, {Sato}, {Sato},
  {Sato}, {Sato}, {Sawada}, {Serlemitsos}, {Seta}, {Shibano}, {Shida},
  {Shimada}, {Shinozaki}, {Shirron}, {Simionescu}, {Simmons}, {Smith},
  {Sneiderman}, {Soong}, {Stawarz}, {Sugawara}, {Sugita}, {Sugita},
  {Szymkowiak}, {Tajima}, {Takahashi}, {Takeda}, {Takei}, {Tamagawa}, {Tamura},
  {Tamura}, {Tanaka}, {Tanaka}, {Tashiro}, {Tawara}, {Terada}, {Terashima},
  {Tombesi}, {Tomida}, {Tsuboi}, {Tsujimoto}, {Tsunemi}, {Tsuru}, {Uchida},
  {Uchiyama}, {Uchiyama}, {Ueda}, {Ueno}, {Uno}, {Urry}, {Ursino}, {de Vries},
  {Wada}, {Watanabe}, {Werner}, {White}, {Yamada}, {Yamada}, {Yamaguchi},
  {Yamasaki}, {Yamauchi}, {Yamauchi}, {Yatsu}, {Yonetoku}, {Yoshida}, \&
  {Yuasa}}]{2012SPIE.8443E..1ZT}
{Takahashi}, T., {Mitsuda}, K., {Kelley}, R., {et~al.} 2012, in Society of
  Photo-Optical Instrumentation Engineers (SPIE) Conference Series, Vol. 8443,
  Society of Photo-Optical Instrumentation Engineers (SPIE) Conference Series

\bibitem[{{Weinberg} {et~al.}(2001){Weinberg}, {Miller}, \&
  {Lamb}}]{2001ApJ...546.1098W}
{Weinberg}, N., {Miller}, M.~C., \& {Lamb}, D.~Q. 2001, \apj, 546, 1098

\bibitem[{{Zink}(2011)}]{2011arXiv1102.5202Z}
{Zink}, B. 2011, ArXiv e-prints

\end{thebibliography}

\end{document}